\documentclass[fleqn,usenatbib]{mnras}

\usepackage{newtxtext,newtxmath}

\usepackage[T1]{fontenc}
\usepackage{ae,aecompl}

\usepackage{graphicx}	
\usepackage{amsmath}	
\usepackage{amssymb}	

\title[O~III offset in dwarf galaxies]{Addressing the [O~III]/H$\beta$ Offset of Dwarf Galaxies in the RESOLVE Survey}

\author[C. T. Richardson et al.]{Chris T. Richardson,$^{1}$\thanks{E-mail: crichardson17@elon.edu (CTR)}
Mugdha S. Polimera, $^{2}$
Sheila J. Kannappan,$^{2}$
\newauthor{Amanda J. Moffett, $^{3}$ and Ashley S. Bittner, $^{4}$}
\\
$^{1}$Physics Department, Elon University, 100 Campus Drive CB 2625, Elon, NC, 27244, USA\\
$^{2}$Department of Physics and Astronomy, University of North Carolina, 141 Chapman Hall CB 3255, Chapel HIll, NC 27599, USA\\
$^{3}$Physics and Astronomy Department, Vanderbilt University, 6301 Stevenson Center, Nashville, TN 37235, USA \\
$^{4}$Department of Civil, Construction, and Environmental Engineering, North Carolina State University, 1205 Burlington Labs, \\
 Raleigh, NC 27695, USA \\
}

\date{Accepted XXX. Received YYY; in original form ZZZ}

\pubyear{2018}

\begin{document}
\label{firstpage}
\pagerange{\pageref{firstpage}--\pageref{lastpage}}
\maketitle

\begin{abstract}

Metal poor dwarf galaxies in the local universe, such as those found in the RESOLVE galaxy survey, often produce high $[$O~III$]$/H$\beta$ ratios close to the star forming demarcation lines of the diagnostic BPT diagram. Modeling the emission from these galaxies at lower metallicities generally underpredicts this line ratio, which is typically attributed to a deficit of photons $>$35 eV. We show that applying a model that includes empirical abundances scaled with metallicity strongly influences the thermal balance in HII regions and preserves the $[$O~III$]$/H$\beta$ offset even in the presence of a harder radiation field generated by interacting binaries. Additional heating mechanisms are more successful in addressing the offset. In accordance with the high sSFR typical of dwarf galaxies in the sample, we demonstrate that cosmic ray heating serves as one mechanism capable of aligning spectral synthesis predictions with observations. We also show that incorporating a range of physical conditions in our modeling can create even better agreement between model calculations and observed emission line ratios. Together these results emphasize that both the hardness of the incident continuum and the variety of physical conditions present in nebular gas clouds must be accurately accounted for prior to drawing conclusions from emission line diagnostic diagrams.

\end{abstract}




\section{Introduction}

The diagnostic diagram [O~III] $\lambda$5007/H$\beta$ vs. [N II] $\lambda$6584/H$\alpha$ (Baldwin, Phillips \& Terlevich (1981); hereafter the BPT diagram) provides a convenient means of classifying galaxies into categories representing the excitation mechanism of nebular gas clouds. In particular, the left hand ``wing" represents a sequence of star forming galaxies that excite clouds via starlight and the right hand ``wing" represents a sequence of active galactic nuclei (AGN) that excite clouds via a combination of thermal and non-thermal sources. Variation in the star forming wing is traditionally interpreted as a change in ionization parameter, $U$, and metallicity, $Z$ (Kewley et al. 2001), or changing distributions in ionizing flux and $Z$ (Richardson et al. 2016, Meskhidze \& Richardson 2017).

At high [O~III]/H$\beta$ along the star forming wing lie galaxies often irregular in morphology and metal poor ($< 0.2 Z_{\odot}$). As such, local galaxies in this region of the BPT diagram are analogous to high-$z$ galaxies, and therefore serve as important examples for tracing galaxy evolution across many epochs. The standard approach to modeling the emission line spectrum of these galaxies involves coupling the spectral energy distribution (SED) outputs from stellar population synthesis (hereafter SPS) codes with spectral synthesis models of the interstellar medium to predict the resulting combined spectrum (e.g. Guktin, Charlot, \& Bruzual 2016; Strom et al. 2017).

Early work showed that the general outcome of this methodology yields a theoretical parameter space typically incapable of matching the observed [O~III]/H$\beta$ of actively star-forming galaxies at low metallicity (Levesque et al. 2010). The predicted ionizing radiation field of secularly evolving (i.e. single) stars cannot explain the observed data at low $Z \approx 0.2 Z_{\odot}$ due to inadequate modeling of Wolf-Rayet (WR) stars. These stars preferentially form at high metallicity in stellar evolution models (Leitherer et al. 1999), but this contradicts observations, in which only extremely metal poor (0.02 $Z_{\odot}$) galaxies show a drop off in WR production (Crowther 2007). Recent modeling has focused on producing more WR stars to match observations, and therefore more photons >35 eV, as a way of alleviating the [O~III]/H$\beta$ offset. One way to generate more WR stars is incorporating differential rotation, which significantly hardens the SED with Starburst99 SPS models (Leitherer et al. 2014). Unfortunately, evolutionary tracks below the 0.4$Z_{\odot}$ have not been generated for Starburst99, limiting use to only 0.4$Z_{\odot}$ and $Z_{\odot}$. 

\begin{figure*}
	\includegraphics[width=\textwidth]{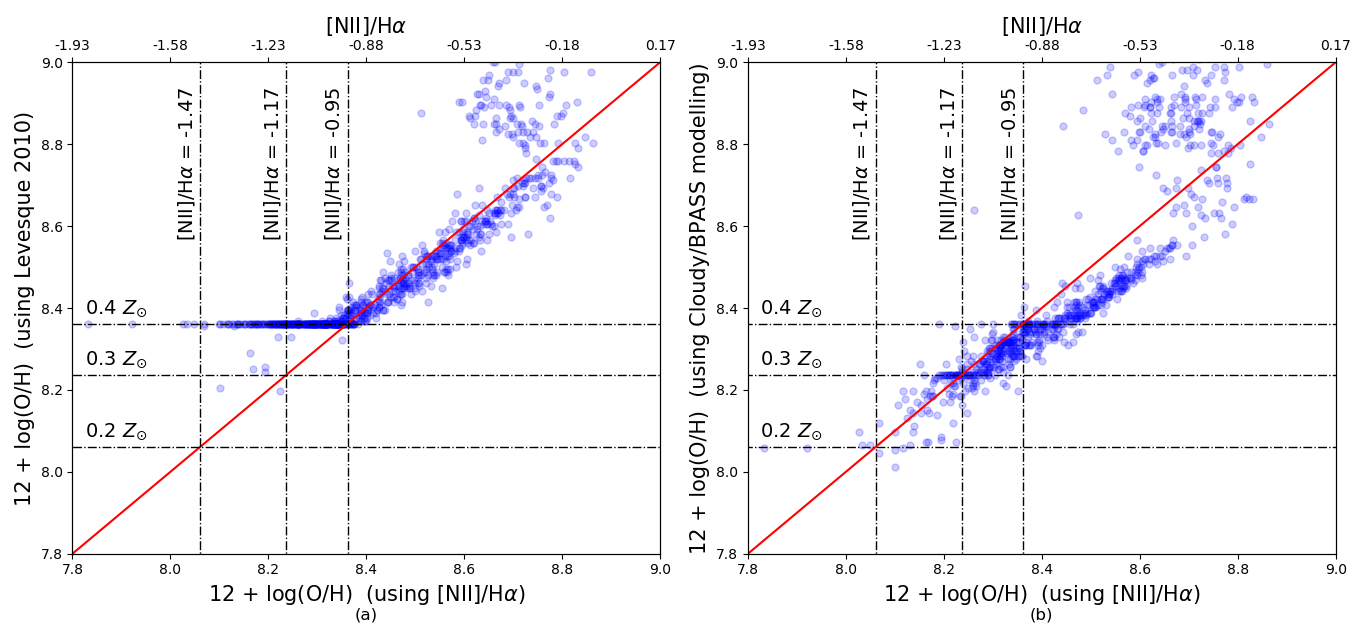}
    \caption{Metallicities for RESOLVE survey galaxies computed with the NebulaBayes code vs.\ metallicities estimated from the [N II]/H$\alpha$ ratio using the empirical calibration of Pettini \& Pagel (2004). Equivalent raw [NII]/H$\alpha$ ratios are shown on the upper axis. The modeling for panel {\it (a)} uses the Levesque et al. (2010) grid while the modeling for panel {\it (b)} uses the Cloudy+BPASS grid described in \S~3.}
    \label{fig:PP04}
\end{figure*}

Another approach to producing photons >35 eV involves expanding stellar evolution to incorporate binaries, which significantly hardens the SED at the low metallicity end of the BPT diagram by including stars entering the WR phase through companion driven mass loss (Stanway, Eldridge, \& Becker 2016). Using the Binary Population and Spectral Synthesis (BPASS) code to create SEDs originating from binary star systems, while coupling this to a simple photoionization model, substantially decreases the [O~III]/H$\beta$ offset between theoretical models and observations (Stanway et al. 2014). These models assume only binary evolution, and while this is unrealistic, it does place useful limits on how much binary evolution can change photoionization model results. As such, this modeling makes basic assumptions about the physical properties of the cloud receiving the ionizing continuum so that any differences in emission line production can be easily attributed to incorporating more realistic SEDs rather than the microphysics in the cloud.

In this paper, we seek to probe many of the simplifying assumptions present in previous work to assess the outcome of alleviating the [O~III]/H$\beta$ offset. We measure self-consistent metallicities using BPASS coupled to the photoionization code Cloudy (Ferland et al. 2017). We assess the effects of using a robust prescription for gas phase chemical evolution, introducing excitation / ionization sources in addition to stellar continua, and using non-uniform distributions of nebular physical conditions. To accomplish this, we use observations from the highly complete REsolved Spectroscopy Of a Local VolumE (RESOLVE) survey (Kannappan \& Wei 2008), which features a significant fraction of star-forming, low metallicity dwarf galaxies, which frequently enter the problematic $Z < 0.4 Z_{\odot}$ regime when using BPASS models.

\section{Observations}

The RESOLVE survey is volume limited in two equatorial footprints between 0.015 $< z <$ 0.023 (4500 $< cz <$  7000 km/s). The survey spans a wide range of environments but is numerically dominated by dwarf galaxies in low-density environments. With its statistically complete design, RESOLVE presents an opportunity to fully understand the necessary gas physics in low metallicity, highly star-forming galaxies. The RESOLVE database also provides supporting data such as stellar masses and star formation rates (Eckert et al.\ 2015, Hood et al.\ 2018).

We start with the 1519 galaxies in RESOLVE Data Release 2 (Eckert et al.\ 2015) that have luminosities above the SDSS selection limit ($-17.33$ in the A semester and $-17.0$ in the B semester) and/or estimated baryonic (stellar+gas) masses above the baryonic mass completeness limit ($10^{9.2}$M$_\odot$ in the A semester and $10^9$M$_\odot$ in the B semester). We then cross-match the RESOLVE coordinates with SDSS DR12 (Alam et al. 2015) to obtain corresponding optical emission-line measurements from the MPA-JHU catalog (Tremonti et al.\ 2004) for our diagnostic diagram analysis. We require a ``reliable'' flag in the spectroscopic catalog and finite, positive fluxes and errors (rejecting flux values $>$10$^5$, which appear to be spurious), yielding 1280 galaxies. For reliable extinction correction and line ratio analysis we further apply a S/N > 3.0 restriction on [O~III] $\lambda$5007, [N II] $\lambda$6584, H$\alpha$, and H$\beta$ and we also remove all galaxies classified as Composite or AGN in the BPT diagram (Kewley et al.\ 2006 following Baldwin et al.\ 1981), yielding an analysis sample of 829 galaxies. Our sample selection introduces a bias against galaxies with high internal extinction, since they will have weak H$\beta$ emission, however these are not generally low-metallicity dwarfs.

We deredden each galaxy's emission lines following Dominguez et al.\ (2013), determining $E(B-V)$ to obtain a dereddened $I($H$\alpha)/I($H$\beta) = 2.86$ as appropriate for Case B recombination lines with an electron temperature $T_e = 10^4$~K and an electron density $n_e = 10^2$ cm$^{-3}$ (Osterbrock \& Ferland 2006). For galaxies with M$_*$>$10^{10}$M$_\odot$, we adopt the Milky Way extinction curve of O'Donnell (1994). For galaxies with M$_*$<$10^{9}$M$_\odot$, we adopt the Small Magellanic Cloud (SMC) extinction curve of Gordon et al.\ (2003); their polynomial fit is poor at optical wavelengths but their data are well fit by a line in this regime, so we use this line for wavelengths $>$3030\AA. At intermediate masses we employ a smoothly varying linear combination of the extinction-corrected fluxes determined with the Milky Way and SMC extinction curves.

Figure~1 assesses the metal-poor nature of our sample galaxies. Evaluations of metallicity depend on the underlying photoionization model. Given that metallicity itself is a tunable parameter in our models that follow, we do not seek a final answer from this preliminary analysis, but rather a baseline calculation that confirms the metal poor nature of dwarf galaxies in RESOLVE, which future work will refine even further. To establish this baseline, we use both an empirical strong line calibration for [N II]/H$\alpha$ (Pettini \& Pagel 2004) and a multiple-line modeling approach with NebulaBayes (Thomas et al.\ 2018, based on Blanc et al.\ 2015). NebulaBayes uses Bayesian inference to derive the metallicity given a set of line fluxes and a photoionization model grid. Our modeling uses H$\alpha$, H$\beta$, [O~III] $\lambda$4959, [O~III] $\lambda$5007, [O I] $\lambda$6300, [N II] $\lambda$6548, [N II] $\lambda$6584, and [S II] $\lambda\lambda$6717, 6731. We consider both a standard photoionization model grid from Levesque et al. (2010) and a custom grid based on Cloudy photoionization modeling (Ferland et al.\ 2017) and BPASS stellar population synthesis modeling (Stanway et al.\ 2016). The finer details of the Cloudy/BPASS grid are given in \S~3 , however unlike in \S~3 , we use a continuous star formation rate of 1 M$_{\odot}$ yr$^{-1}$ and an age of 40 Myr, which is when the ionizing continuum reached steady state for all metallicities. We adjust the Levesque grid from Z$_\odot$ = 8.66 to 8.76 to match the other two sources of metallicities. As seen in Fig.~1, the Levesque grid hits a floor at low metallicity, but both the empirical [N II]/H$\alpha$ calibration and our Cloudy/BPASS grid confirm a significant population of low-metallicity galaxies ($\sim$43\%\ at Z $<$ 0.4 Z$_\odot$ and $\sim$14\%\ at Z $<$ 0.3 Z$_\odot$ using the Cloudy/BPASS grid). In the remainder of this paper, we focus on the 139 galaxies with log([NII]/H$\alpha$) $<$ $-1$, representing 17\%\ of the original 829 galaxy emission-line sample.

\section{Simulations}

\begin{figure*}
	\includegraphics{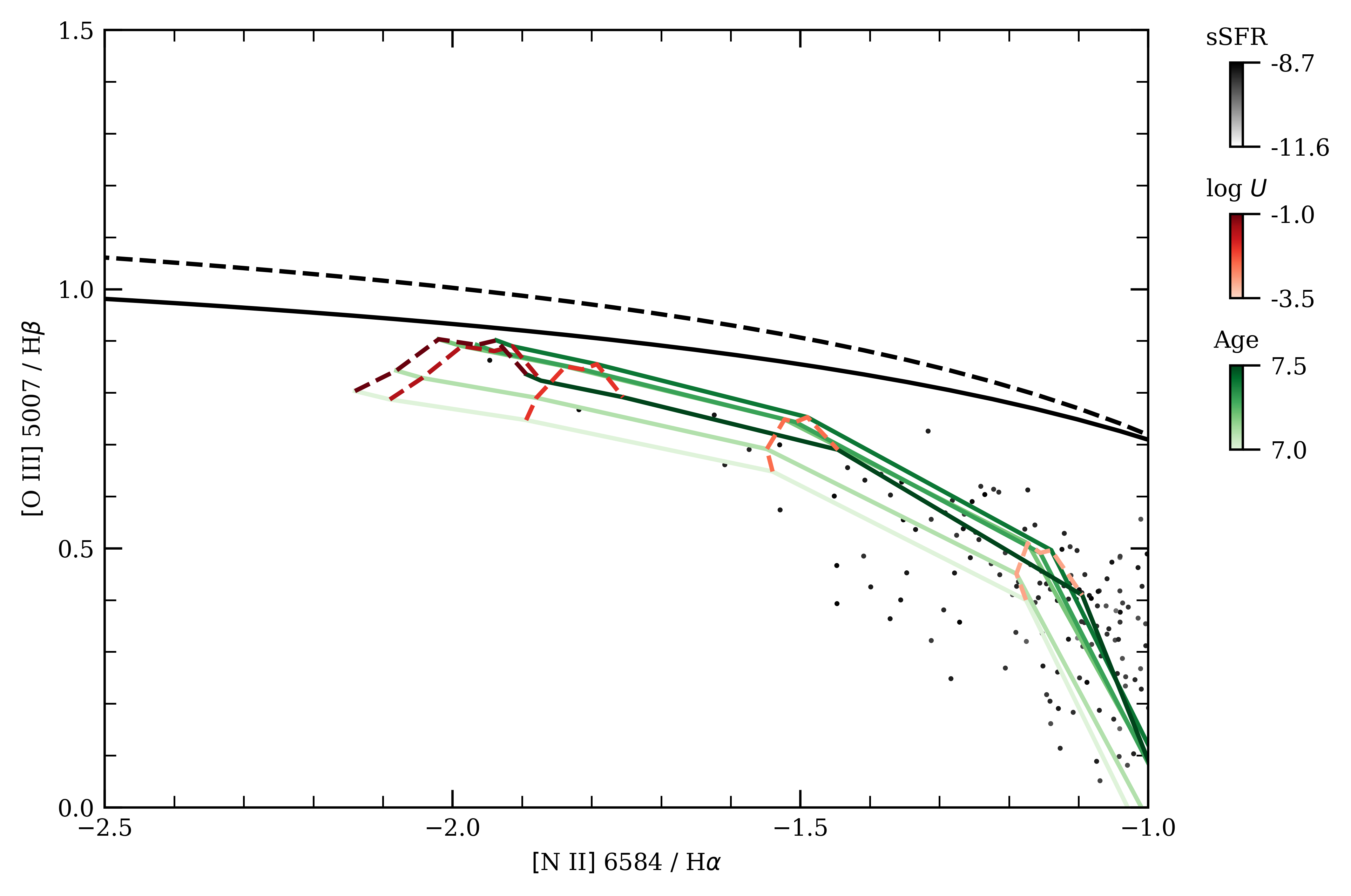}
    \caption{BPT diagram at $0.3 Z_{\odot}$ using the BPASS stellar population synthesis code. A significant [O~III]/H$\beta$ offset is present at [N II]/H$\alpha$ = -1.17. SSFRs are based on SFRs from Hood et al. (2018) and stellar masses from Eckert et al. (2015).}
    \label{fig:BPASS-0_3Z}
\end{figure*}

\begin{figure*}
	\includegraphics{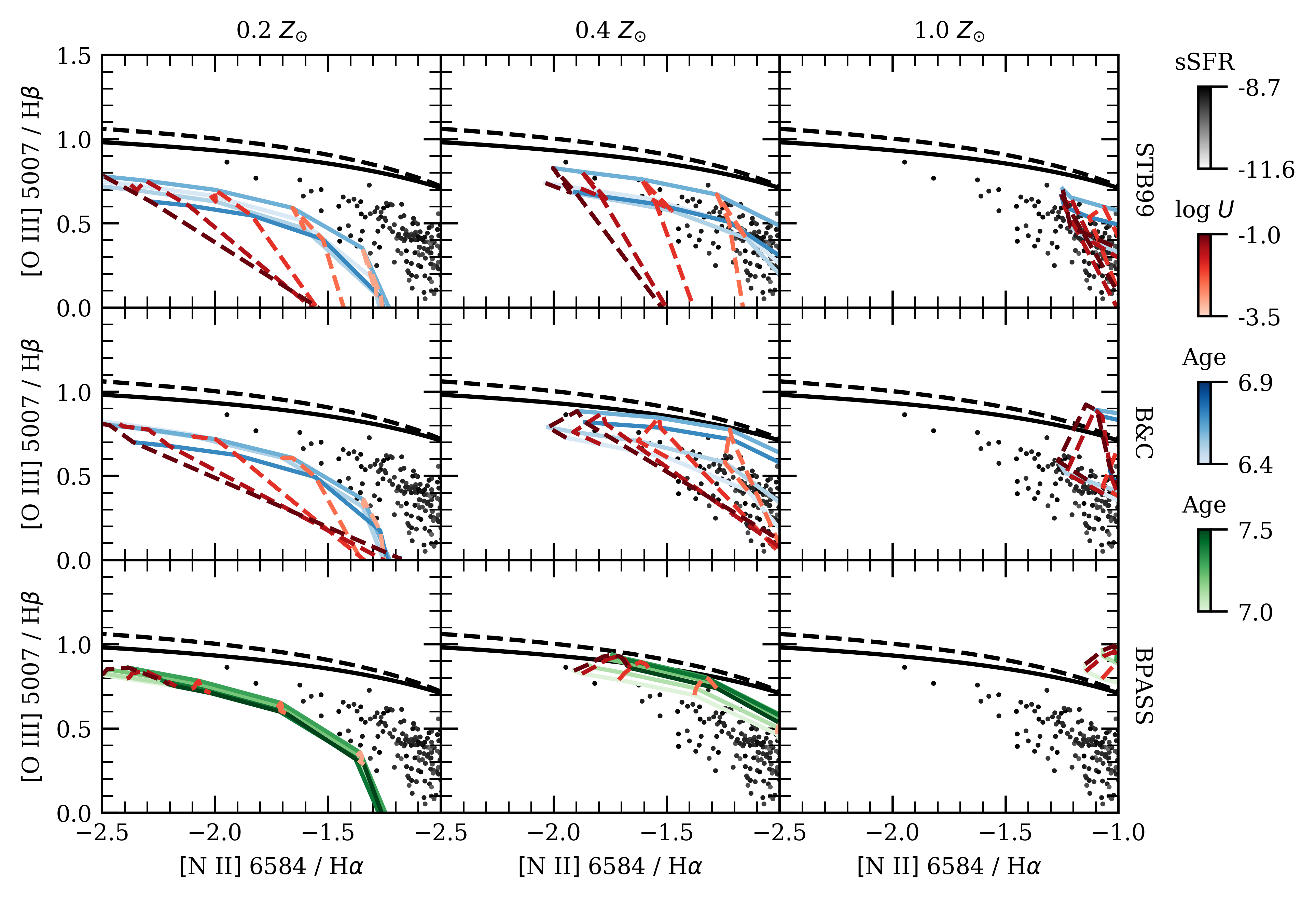}
    \caption{BPT diagrams with columns corresponding to metallicity and rows corresponding to different stellar population synthesis models. In each case $Z_{\mathrm{stellar}} = Z_{\mathrm{gas}}$, which is scaled according to Nicholls et al. (2017). The lowest $Z$ column shows the [O~III]/H$\beta$ offset is present for all three models.}
    \label{fig:secC+N}
\end{figure*}

\textit{Stellar SEDs}: We use three SPS models to generate the radiation field incident on the gaseous clouds in the star forming regions: Starburst99 (Leitherer et al. 1999, hereafter STB99), Bruzual \& Charlot (2003, hereafter B\&C), and BPASS v2.0 (Stanway, Eldridge, \& Becker 2016). All of the continua generated using the codes assume a single stellar population with a fixed mass of $10^6$ M$_{\odot}$ at metallicities of 0.2 Z$_{\odot}$, 0.4 $Z_{\odot}$, and 1.0 $Z_{\odot}$ for STB99 and B\&C, while BPASS has an additional metallicity at 0.3 Z$_{\odot}$. 

For STB99 models, we adopt a Kroupa (2001) initial mass function (IMF) with exponents of 1.3 and 2.3 for the broken power law over the ranges of 0.1 M$_{\odot}$ < M < 0.5 M$_{\odot}$ and 0.5 M$_{\odot}$ < M$_{\odot}$ < 100.0 M$_{\odot}$. We use the Padova AGB evolutionary track (Bressan et al. 1993), which extends down to 0.2 Z$_{\odot}$ unlike the newer Geneva tracks that incorporate differential stellar rotation. All of the other STB99 parameters are left at their default values.

For B\&C models, we adopt a Chabrier (2003) IMF with the empirically fitted form as in Chabrier (2003) for 0.1 M$_{\odot}$ < M < M$_{\odot}$ and a power law index of 2.3 for M$_{\odot}$ < M < 100 M$_{\odot}$.

For BPASS models, we adopt a Kroupa initial mass function (IMF) with exponents of 1.3 and 2.35 for the broken power law over the ranges of 0.1 M$_{\odot}$ < M < 0.5 M$_{\odot}$ and 0.5 M$_{\odot}$< M < 300 M$_{\odot}$. We only consider binary evolution without any secularly evolving stars. 

The limitations of what is freely available for each code did not allow us to provide the exact same IMF in each case, however in the limits investigated here, the Kroupa and Chabrier forms are very similar. The most distinguishing difference comes from the upper mass limit of 300 M$_{\odot}$ in the BPASS models. We select this upper bound, which is higher than the 100 M$_{\odot}$ bound for STB99 and B\&C, to explore the maximum flux that can conceivably be emitted at >35~eV using binary evolution. Similarly, in order to maximize high-energy photons, our stellar age ranges were all stepped through in 1 Myr increments and included the age at which each SPS model generates its maximum ionizing continuum.

\begin{figure*}
	\includegraphics{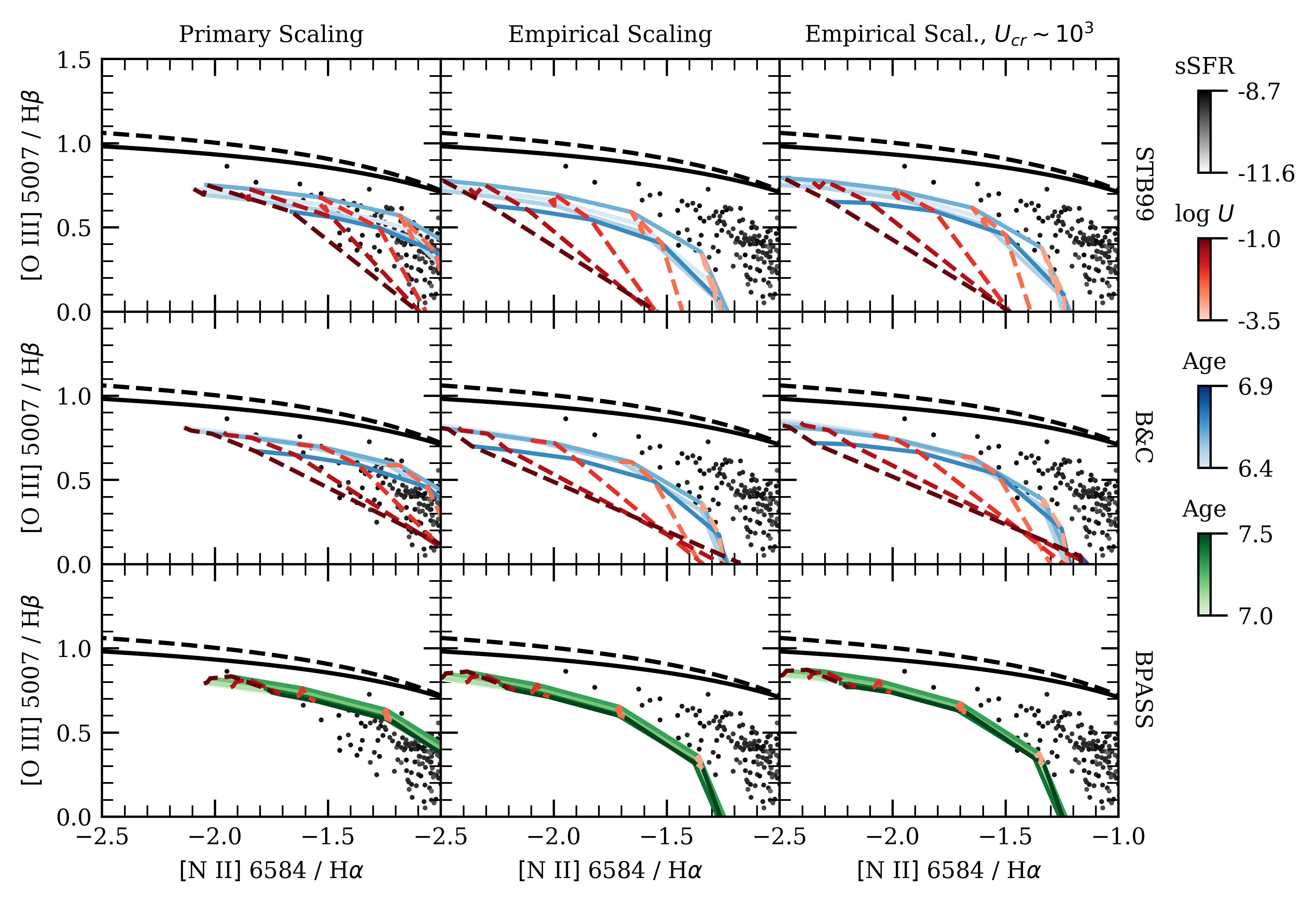}
    \caption{BPT diagrams at $0.2 Z_{\odot}$ with columns corresponding to assumed chemical evolution and presence of cosmic rays, and rows corresponding to different stellar population synthesis models. The dashed lines with a red color map refer to simulations with constant $U$, while the solid lines with either blue or green color maps refer to simulations with constant starburst age.  Assuming that all elements scale linearly with $Z$ eliminates the offset, however several elements are known to deviate from this relationship at $0.2 Z_{\odot}$. Additional heating through cosmic rays does not appreciably reduce the offset after assuming a more robust empirical scaling for all elements.}
    \label{fig:secC+N-cr}
\end{figure*}

\textit{Cloud Properties}: We use Cloudy v17.01 (Ferland et al. 2017) to handle the necessary microphysics in the cloud and predict the emitted spectrum. For our solar abundance set, we adopt the Galactic Concordance abundances given Nicholls et al. (2017), which are based off of Lodders, Palme \& Gail (2009), Nieva \& Przybilla (2012), Grevesse et al. (2015), and Scott et al. (2015a,b). In this scale, the solar oxygen abundance is 12+log(O/H) = 8.76. The chemical evolution model in Nicholls et al. (2017) provides a robust method for uniquely scaling all elements over a wide range of metallicities (e.g., primary and secondary nucleosynthesis for N and C) according to the scaling parameter $\zeta_\mathrm{O}$, which we will refer to as the metallicity even though it is not strictly a mass fraction (see Hamann et al. 2002 for a discussion). For simplicity, when using this method of scaling we assume that gas-phase metallicities match the stellar metallicities of the chosen SPS model. We note that our abundance scaling only applies to the nebular composition, and not to the stellar composition, but scaling the stellar component similarly would have a relatively minor effect.

We use the grain composition of the Orion Nebula for our models by incorporating silicate and graphite grains with sizes according to Baldwin et al. (1991) and polycyclic aromatic hydrocarbons (PAHs) according to Abel et al. (2008). Depletion factors due to grains vary widely at a given metallicity and as a function of metallicity (De Cia et al. 2016), but for simplicity we do not attempt to account for this. We select the depletion factors in Groves et al. (2004), except for the nitrogen depletion factor of 0.11 dex, which is the average of the Groves et al. (2004a) and default Cloudy nitrogen depletion values and agrees with Jenkins (2014). We scale our grain abundances linearly with metallicity. Given the rarity of PAHs prior to the ionization front (Sellgren et al. 1990), we also scale our PAH abundance with $n($H$^0)$/n(H) throughout the entire cloud. We assume that constant total pressure holds in each cloud while enforcing a boundary condition at $n_{\mathrm{e}}/n_{\mathrm{H}}=0.1$.

\subsection{Single Cloud Models}

Our single cloud models assume a single density and ionization parameter are common to all clouds within a given galaxy. This simple methodology has been successful in spanning most of the galaxies on the BPT diagram (e.g., Kewley et al. 2013) and therefore forms the starting point for our modeling. The median density in the RESOLVE survey as derived from the [S II] ratio with the calibration in Proxauf, \"{O}ttl, and Kimeswenger (2014) gives log $n_{\mathrm{H}}$ = 2.0, which we set for all our models in our initial set of simulations. The ionization parameter, $U$, is defined by:

\begin{equation}
    U = \frac{\phi_{\mathrm{H}}}{n_{\mathrm{H}} c}
	\label{eq:U}
\end{equation}

\noindent where $\phi_{\mathrm{H}}$ is the hydrogen ionizing flux. We run simulations using six ionization parameters log $U$ = -3.5, -3.0, -2.5, -2.0, -1.5, and -1.0. 

Given that the [O~III]/H$\beta$ offset occurs at lower metallicities and Fig. 1b shows that $\sim$13\% of the entire RESOLVE sample has Z $<$ 0.3~Z$_\odot$, we selected the BPASS model at 0.3 Z$_\odot$ as our initial proof of concept model. Fig. 2 shows the results of this model with demarcation curves from Kewley et al. (2001) and Kauffmann et al. (2003). At [N~II]/H$\alpha$ = -1.17, which corresponds to 0.3~Z$_\odot$ as seen in Fig. 1, several galaxies show [O~III]/H$\beta$ emission $\sim$0.1 dex greater than what our models can predict. The galaxies that sit below our models predictions for [O~III]/H$\beta$ can be fit by including ages either below 10 Myr or above $\sim$30 Myr, when the ionizing continuum significantly weakens, but we have not included these ages to maintain consistency with Figs. 3-5. Similarly, the galaxies with enhanced [O~III]/H$\beta$ emission relative to our models in Fig. 2 at [N~II]/H$\alpha$ < -1.17 can be fit with models at 0.4~Z$_\odot$, as shown in Fig. 3.

Unfortunately, only BPASS has a model at 0.3 Z$_\odot$, which strongly limits our analysis at this metallicity. However, the metallicity predictions using NebulaBayes are rather sensitive to the underlying photoionization model supplied to the code, with certain models yielding substantially more galaxies at 0.2 Z$_\odot$ rather than 0.3 Z$_\odot$, and other models yielding substantially more galaxies at 0.4 Z$_\odot$ than 0.3 Z$_\odot$ (e.g. Levesque et al. 2010). Elaborating on this will be the subject of a forthcoming paper (Moffett et al., in prep), but for now we go forth with STB99, B\&C, and BPASS models at 0.4 Z$_\odot$ and 0.2 Z$_\odot$ for galaxies with [N II]/H$\alpha$ $\leq$ -1.17, because these metallicity values are common to all SPS models and galaxies at 0.2 Z$_\odot$ also show a significant [O~III]/H$\beta$ offset.

\begin{figure*}
	\includegraphics{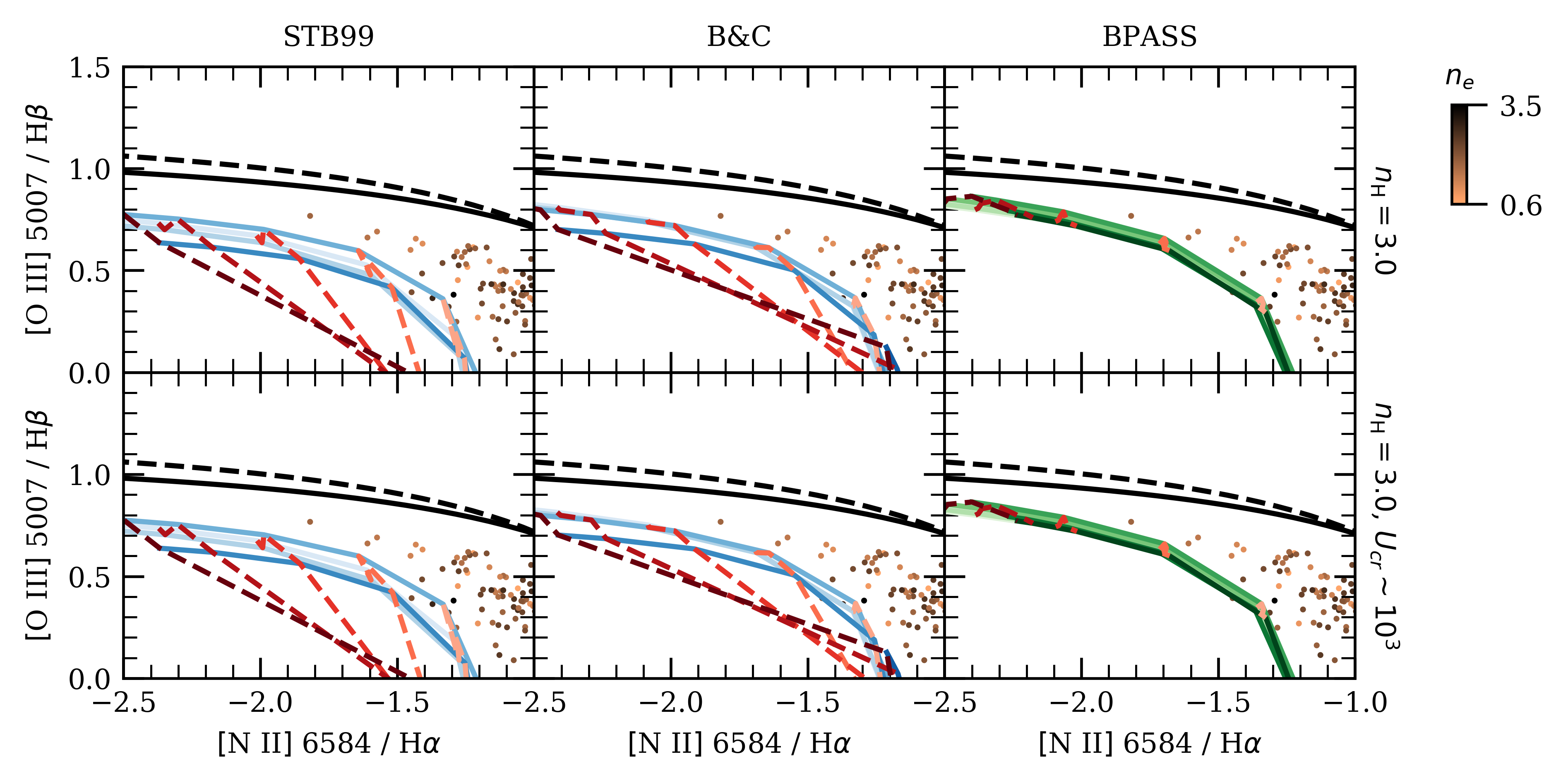}
    \caption{BPT diagrams at $0.2 Z_{\odot}$ with rows corresponding to different stellar population synthesis models and columns corresponding to initial $n_{\mathrm{H}}$ and presence of cosmic rays. Increasing the initial density to the maximum calculated from observations marginally reduces the offset, as with introducing additional cosmic rays for additional heating.}
    \label{fig:secC+N-nH-cr}
\end{figure*}

Fig. 3 compares observations to our models as a function of Z and SPS model across the BPT diagram. The general trends present when varying ionization parameter and stellar age are similar in each of the diagrams. The trend with ionization parameter is similar for all models with [N II] predominantly affected at high $U$ and [O~III] predominantly affected at low $U$, when the fraction of O$^{++}$ significantly drops off. The starburst age, and therefore the ionizing continuum, produces a $\sim$0.2 dex variation in [O~III]/H$\beta$ over a 3-4 Myr range before significantly dropping [O~III] emission production as O \& B stars die off. The notable exception to this trend is with the BPASS models, which sustain a hard continuum for longer periods of time at all metallicities but span a smaller range in [O~III]/H$\beta$ at 0.2 Z$_{\odot}$.

All models at 0.2 Z$_{\odot}$ emphasize the 0.2-0.3 dex discrepancy between [O~III]/H$\beta$ predictions and observations. In contrast, all models at 0.4 Z$_{\odot}$ can eliminate the [O~III]/H$\beta$ offset, highlighting the need for model grids that bridge between 0.2~Z$_{\odot}$ and 0.4~Z$_{\odot}$, as seen in Fig. 1b, but not Fig. 1a. 

Indeed, galaxies with predicted metallicities of 0.2-0.3 Z$_{\odot}$ (12+log(O/H) $\approx$ 8.1) with BPASS or the empirically PP04 calibration are predicted to have an approximate mean metallicity of 0.4 Z$_{\odot}$ using the Levesque et al. (2010) grid, although it includes a grid point at 0.2~Z$_{\odot}$. Our results maximize the possible role of binaries because our BPASS models assume a pure binary population. However, even secular BPASS models include differential rotation, which already reduces traditionally derived metallicities and enhances [O~III]/H$\beta$ relative to previous methods. Further investigation of the low metallicities found with BPASS is forthcoming (Moffett et al., in prep).

Finally, as expected, solar metallicity models never approach the upper left hand side of the star forming BPT wing due to the metallicity sensitive [N II]/H$\alpha$ ratio. The BPASS models extend well beyond the demarcation curves determined with older models, confirming the subsolar gas metallicities across much of the star forming wing of the BPT diagram.

While the models using BPASS reduce the deficit at 0.2~Z$_{\odot}$ more than those with B\&C or STB99, BPASS does not eliminate the need to improve our current predictions from emission line modeling at 0.2 Z$_{\odot}$. We have tried assuming that the stellar metallicity and gas metallicity are dissimilar, for example from recently accreted gas. This also does not solve the problem, with low gas metallicity simulations still showing a significant offset.

A detailed analysis of mixing these metallicities is beyond the scope this work; however, focusing on the exact nature of scaling abundances as function of $Z$ in a way consistent with chemical evolution helps contextualize work using different prescriptions (e.g. Kewley et al. 2013, Eldridge et al. 2017). In the first two columns of Fig. 4, we highlight two different methods for accounting for chemical evolution at 0.2 Z$_{\odot}$. The first column assumes all elements obey a primary scaling, such that [X/H] $\propto$ Z where X is a gas phase element, while the second column is shown for reference using the empirical prescription described in \S 3. For reference, in the empirical scaling, this translates to the N/O ratio changing by a factor of 0.338 at 0.2 Z$_{\odot}$ relative to the primary model.

In the primary only case, all of the [O~III]/H$\beta$ ratios are increased substantially, independent of SPS model, with BPASS slightly spanning a greater range if we extend our analysis to a greater age range. Unfortunately, the main caveat is that many elements do not obey a simple scaling at this metallicity, for example the secondary nucleosynthesis components for N and C (Nicholls et al. 2017). These results emphasize that inaccurate treatment of the gas physics can hide a real discrepancy between observations and model predictions. Thus, thermal balance plays an important role in understanding how to reduce the [O~III]/H$\beta$ offset.
	
The high sSFR rates shown in RESOLVE galaxies in the upper left of the star forming wing (grayscale, Figs. 2 \& 3) suggest several other potential factors that could affect physical processes, including an increased fraction of x-ray binaries (Mineo, Gilfanov, \& Sunyaev 2012), turbulent flow from infalling gas (Kraemer, Bottorff, \& Crenshaw 2007; Gray \& Scannapieco 2017), and cosmic ray heating (Meijerink et al. 2011; Papadopoulos et al. 2011). As an exploration, we run simulations that incorporate cosmic ray heating as described in Richardson et al. (2013). We adopt a cosmic ray density of $U_\mathrm{cr} = 10^3$ relative to a galactic background rate of $\xi = 2.0 \times 10^{-16} ~ \mathrm{s}^{-1}$ (Indriolo et al. 2007), which puts our increased rate within reason of starburst galaxies (Suchkov et al. 1993; Acero et al. 2009; Pagoline \& Abrahams 2012). 

While increased heating decreases the [O~III]/H$\beta$ offset by $\sim$0.05 dex, the fit is still unsatisfactory. Until this point, we have assumed conditions typical for RESOLVE dwarf galaxies; however, the few galaxies that display high [O~III]/H$\beta$ ratios might also possess conditions atypical of the entire sample. In particular, enhanced hydrogen densities increase collisional excitation and therefore strengthen relative [O~III] line emission (Xiao, Stanway, \& Eldridge 2018). Using the empirical fit from Proxauf, \"{O}ttl, and Kimeswenger (2014), we find a maximum of $n_\mathrm{e} = 10^{3.5}$ cm$^{-3}$ for the RESOLVE survey after requiring S/N $>$ 3.0 for [S II] $\lambda\lambda$6716, 6731.

\begin{figure*}
	\includegraphics{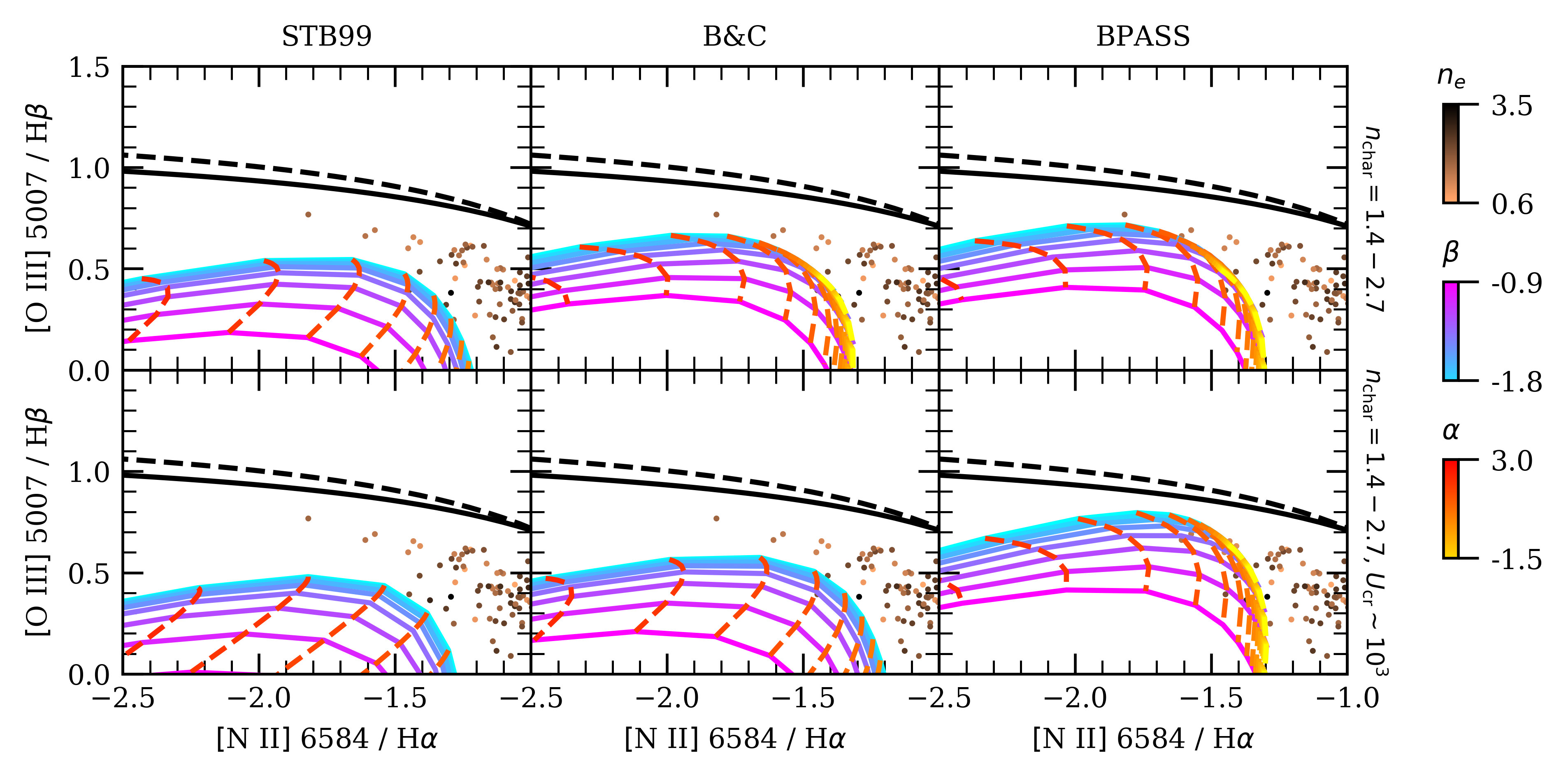}
    \caption{BPT diagrams at $0.2 Z_{\odot}$ for an ensemble cloud model with rows corresponding to different SPS models and columns corresponding to characteristic density $n_{\mathrm{H}}$ and presence of cosmic rays. Adopting a more sophisticated model with an ensemble of clouds does not reduce the [O~III]/H$\beta$ offset, however also introducing additional heating via cosmic rays essentially eliminates the offset regardless of the stellar population model.}
    \label{fig:spop_0_2Z-cr}
\end{figure*}

Fig. 5 shows the distribution of $n_\mathrm{e}$ for our sample in each panel. Galaxies exhibiting the offset generally do not have a higher calculated $n_\mathrm{e}$, but clouds with enhanced $n_\mathrm{e}$ could still exist within those galaxies. The gas clouds that generate the bulk of the [S II] emission in our observations could differ from those that constitute the bulk of the [O~III] emission. As such, Fig. 5 displays our models with hydrogen density increased to $n_\mathrm{H} = 10^{3}$  cm$^{-3}$, while the bottom row also includes additional cosmic ray heating. The impact is again quite minimal in reducing the offset between observations and predicted emission line ratios.

\subsection{Distribution Cloud Models}
\label{sec:dist_models} 

Thus far, our models have assumed a single ionization parameter and density can characterize the properties of an entire galaxy. This implies that all H II regions within a galaxy are the same size. In reality, H II regions vary drastically in size with a variety of physical conditions. We account for this fact by running suites of models varying $\phi_\mathrm{H}$ and $n_\mathrm{H}$ to form a parameter space that captures the peak emission for most emission lines in the optical (Meskhidze \& Richardson 2017). This approach addresses the restriction mentioned earlier, where clouds with different conditions are responsible for generating the observed emission lines strengths. We then perform a weighted integration over the entire parameter space, using power laws for simplicity, according to

\begin{equation}
    L_{\mathrm{ion}} = \int \int F(\phi_{\mathrm{H}}, n_{\mathrm{H}}) ~ \phi_{\mathrm{H}}^{\alpha} n_{\mathrm{H}}^{\beta} ~ d n_{\mathrm{H}} d \phi_{\mathrm{H}}
	\label{eq:loc_int}
\end{equation}

\noindent where $\alpha$ and $\beta$ are the integration weightings for $\phi_\mathrm{H}$ and $n_\mathrm{H}$, respectively (Richardson et al. 2016). Adjusting these weightings places more or less emphasis on certain parts of the $\phi_\mathrm{H}$-$n_\mathrm{H}$ parameter space, effectively indicating that certain H II regions will dominate observations where the aperture only measures global properties. Similarly, one can define a characteristic density,

\begin{equation}
    n_\mathrm{char} = \frac{\Sigma^N_{i=0} n_\mathrm{H} W_\lambda \phi_\mathrm{H}^\alpha n_\mathrm{H}^\beta}{\Sigma^N_{i=0} W_\lambda \phi_\mathrm{H}^\alpha n_\mathrm{H}^\beta}
	\label{eq:n_char}
\end{equation}

\noindent weighted by  $\alpha$, $\beta$, and emission line equivalent width, $W_\lambda$, which represents the density that would be determined from spatially unresolved, and thus biased, observations. Our parameter space spans $n_\mathrm{H} = 10^{1}-10^{6}$  cm$^{-3}$ and $\phi_\mathrm{H} = 10^{8.9}-10^{16.9}$ erg s$^{-1}$ cm$^{-3}$ while varying $-1.8 < \beta < -0.9$ and $-3.5 < \alpha <-1.5$.

For these photoionization models, we select the age for each SPS model corresponding to the hardest continuum at 0.2 Z$_{\odot}$. This translates to 4 Myr for both B\&C and STB99, but 20 Myr for BPASS due to the delayed effects associated with binary evolution. Fig. 6 displays our models with an ensemble cloud distribution. The top row shows that the ensemble cloud model using BPASS is more successful than any of the single cloud models using BPASS since it matches more of the galaxies that show enhanced [O~III]/H$\beta$ emission. While less of a concern at high ionization, the ensemble cloud model also does not show the same level of degeneracy in free parameters as the single cloud models.

Introducing an ensemble of clouds does not fully address the discrepancy between our models and observations, so we move forward by incorporating enhanced cosmic ray excitation. The bottom row of Fig. 6 displays the result, which essentially eliminates the [O~III]/H$\beta$ offset with BPASS. The B\&C and STB99 models still underpredict [O~III]/H$\beta$ as with previous simulations.

The characteristic densities found for the $\alpha$ and $\beta$ values typical of galaxies with large [O~III]/H$\beta$ offsets range from $n_\mathrm{char} = 1.4$ to $n_\mathrm{char} = 2.7$, which are at the higher end of the RESOLVE density distribution. There is not a strict correlation between our models with higher $n_\mathrm{char}$ and galaxies that indicate higher $n_\mathrm{e}$, and Fig. 5 shows very large scatter in $n_\mathrm{e}$ at low metallicity. A few reasons could explain this lack of relationship. First, we assumed simple power laws for our distribution function, while the true distribution could be Gaussian or even fractal in nature (Bottorff et al. 2001).

Second, we assumed a single stellar age and metallicity. As discussed in \S 3.1, age plays an important role in dictating the predicted [O~III]/H$\beta$, and therefore several combinations of $n_\mathrm{char}$ and stellar age could produce similar results. Similarly, one can also envision distributions of metallicity within a single galaxy instead of a uniformly singular value as our models assume. In this case, a global metallicity would be deduced from unresolved observations when in reality higher $Z$ gas clouds (see \S 3.1) could be responsible for [O~III]/H$\beta$ production while low $Z$ gas clouds produce the bulk of the [N II]/H$\alpha$ emission. Indeed, multiple degeneracies likely exist in reducing the [O~III]/H$\beta$ offset and Bayesian approaches show promise in breaking them (Thomas et al. 2018).

\section{Implications and Conclusions}

Reflecting upon the sensitivity of our emission line modeling to various physical parameters, we can compile a rough ranking of how much they affect the [O~III]/H$\beta$ offset from the least influential to the most influential:
\itemize{\item \textit{Ionization Parameter}: Galaxies with large [O~III]/H$\beta$ are relatively insensitive to $U$ with much more noticeable effects along the lower portion of the star-forming wing.
\item \textit{Cosmic Ray Excitation}: Up to 0.05-0.1 dex of the [O~III]/H$\beta$ offset can be attributed to increased cosmic ray flux due to high rates of star formation.
\item \textit{SPS Model}: Surprisingly, the choice in SPS model as it relates to the maximum ionizing flux changes [O~III]/H$\beta$ offset at most by 0.05 dex, despite the harder radiation field from binaries. The effect is likely even smaller since including a IMF upper limit of 300 M$_{\odot}$ and purely binary stellar systems are unrealistic assumptions. The greatest difference arises from the ability of BPASS models to produce harder SEDs for a greater range of stellar ages making the necessary physical conditions less fine tuned.
\item \textit{Nebular Selection Effects}: Including many gas clouds with a wide range of physical conditions can eliminate almost 0.1 dex of the offset.
\item \textit{SPS Age}: Stellar age can decrease the [O~III]/H$\beta$ offset by 0.2 dex for SEDs capable of making substantial amounts of O$^{++}$. Thus, the biggest asset of BPASS is the ability to sustain a hard continuum for longer periods of time.
\item \textit{Abundances / Metallicity}: Differences between empirical and primary scaling of abundances results in 0.1 dex difference in [O~III]/H$\beta$, while scaling metallicity from 0.2 Z$_{\odot}$ to 0.4 Z$_{\odot}$ reduces the offset by 0.2 dex.}\\

We determined that the method in scaling abundances with metallicity fundamentally affects the predictions from our photoionization modeling, in particular, accounting for secondary nucleosynthesis in nitrogen. This sensitivity brings up several caveats related to abundances that serve as motivation for future work. First, there still remains little consensus on a ``standard" abundance set in nebular astrophysics. Solar abundances (Grevesse et al. 2010), B-star abundances (Nieva \& Przybilla 2012), and H II region abundances (Baldwin et al. 1991) agree to a rough extent, but we have shown here that relatively small changes can influence attributing the [O~III]/H$\beta$ offset to a hard photon problem rather than a thermal balance problem. In a similar vein, we scaled our grain abundances linearly with gas metallicity, although Remy-Ruyer et al. (2014) shows for a given metallicity the D/G ratio can vary as much as 2 dex depending on stellar population age. Such changes require knowledge of how metal depletion varies at a fixed metallicity and as a function of metallicity, both of which are poorly known but improving (De Cia et al. 2016). All together, improving the determination of these attributes would contribute to better understanding of the [O~III]/H$\beta$ offset.

Our best models result from an assumed power law distribution in ionizing flux and density within each star-forming galaxy, assumed to have uniform metallicity. We chose this distribution for simplicity, and based off of previous work, although measuring the actual distributions of density and metallicity would help constrain our models. RESOLVE's 3D spectroscopy will serve as the foundation for improved photoionization simulations in this regard. Similarly, our best models incorporate cosmic ray excitation under the presumption that each gas cloud sees that same cosmic ray flux. While this is unlikely, our value is below the maximum calculated value in starburst galaxies, so one can imagine that an ensemble of gas clouds seeing different fluxes could average out to our adopted value.

Our results imply that star-forming galaxies with high [O~III]/H$\beta$ ratios can result from low metallicity gaseous clouds with a range of ionizing fluxes and densities that also experience excitation from an increased cosmic ray density due to high sSFR. The most successful model in this regard comes from using the BPASS code, but also requires the ages of the clusters dominating the emission line observations to fall within the range that produces a hard ionizing continuum. This work represents progress in understanding the degeneracies present when modeling galaxies across the BPT diagram and serves as an incentive for increased comprehension of abundance and dust evolution with cosmic time.

\section*{Acknowledgments}

CR gratefully acknowledges the support of the Elon University FR\&D committee and the Extreme Science and Engineering Discovery Environment (XSEDE), which is supported by National Science Foundation grant number ACI-1548562. This work used the XSEDE resource Comet at the San Diego Supercomputing Center through allocation TG-AST140040. AM acknowledges funding from the Vanderbilt University Stevenson Postdoctoral Fellowship. We thank Vianney Lebouteiller, Peter van Hoof, and Gary Ferland for helpful discussions and comments.







\bsp	
\label{lastpage}
\end{document}